AUTHORS:

Erik Kusch[1], Richard Davy[2]

AFFILIATIONS:

[1]Center for Biodiversity Dynamics in a Changing World (BIOCHANGE), Section for Ecoinformatics & Biodiversity, Department of Biology, Arhus University

[2]Nansen Environmental and Remote Sensing Center, Thormohlensgate 47, Bergen, Norway

CORRESPONDING AUTHOR:

Erik Kusch; erik.kusch@bio.au.dk

Department of Biology

Section for Ecoinformatics & Biodiversity

Center for Biodiversity Dynamics in a Changing World (BIOCHANGE)

Arhus University

**ABSTRACT**

1. Advances in climate science have rendered obsolete the gridded observation data sets commonly used in macroecological analyses. Novel climate reanalysis products outperform legacy data products in accuracy, temporal resolution, and provision of uncertainty metrics. Consequently, there is an urgent need to develop a workflow through which to integrate these improved data into biological analyses.

2. The ERA5 product family (ERA5 and ERA5-Land) are the latest and most advanced global reanalysis products created by the European Center for Medium-range Weather Forecasting (ECMWF). These data products offer up to 83 essential climate variables (ECVs) at hourly intervals for the time-period of 1981 to today with preliminary back-extensions being available for 1950-1981. Spatial resolutions range from 30x30km (ERA5) to 11x11km (ERA5-Land) and can be statistically downscaled to study-requirements at finer spatial resolutions. Kriging is one such method to interpolate data to finer resolutions and has the advantages that one can leverage additional covariate information and obtain the uncertainty associated with the downscaling.

3. The KrigR R Package enables users to (1) download ERA5(-Land) climate reanalysis data for a user-specified region, and time-period, (2) aggregate these climate products to desired temporal resolutions and metrics, (3) acquire topographical co-variates, and (4) statistically downscale spatial data to a user-specified resolution using co-variate data via kriging. KrigR can execute all these tasks in a single function call, thus enabling the user to obtain any of 83 (ERA5) / 50 (ERA5-Land) climate variables at high spatial and temporal resolution with a single R-command. Additionally, KrigR contains functionality for computation of bioclimatic variables for use in macroecological studies.

4. This R package provides an easy-to-implement workflow for implementation of state-of-the-art climate data into biological analyses while avoiding issues of storage limitations at high temporal and spatial resolutions by providing data according to user-needs rather than in global data sets. Consequently, KrigR provides a toolbox to obtain a wide range of tailored climate data at unprecedented combinations of high temporal and spatial resolutions thus enabling the use of world-leading climate data in the R-environment.

1. **INTRODUCTION**

*CLIMATE DATA NEEDS IN THE 21ST CENTURY*

With the onset of the Anthropocene, the numerous fields of study that investigate the effects of climate change require spatially and temporally consistent climate data at high spatial and temporal resolutions (Hewitt, Stone, & Tait, 2017; Bjorkman et al., 2018; Trisos, Merow, & Pigot, 2020). In response to this need, an ever-growing number of climate datasets have been created (Fick & Hijmans, 2017; Karger et al., 2017; Abatzoglou, Dobrowski, Parks, & Hegewisch, 2018; Beyer, Krapp, & Manica, 2020; Navarro-Racines, Tarapues, Thornton, Jarvis, & Ramirez-Villegas, 2020) making use of observations, reanalysis products, climate model outputs, or some combination thereof. Historically, efforts of climate data product creation for use in macroecological analyses have prioritised spatial resolution over temporal resolution in-line with the widely accepted notion of small-scale processes affecting large-scale patterns (Briscoe et al., 2019; Rapacciuolo & Blois, 2019). This has resulted in climate products at spatial resolutions of up to 30 arcseconds (~900m) which are typically available at monthly or climatological-mean temporal resolutions. In view of climate-change effects on microclimatic processes as well as the changing frequency and intensity of climatic extremes, this emphasis on spatial rather than temporal resolution has led to a decreased ability to identify extreme events and their consequences (Maclean, 2019). Consequently, there is a pressing necessity for the development and dissemination of climate data products that offer data at high spatial and temporal resolutions.

Accurate representation of environmental conditions is facilitated not just through high spatial and temporal resolutions of climate data, but also through a wide range of climate variables. Contemporary studies of environmental drivers of biological patterns and processes have focused on a wide range of environmental variables including (1) water-availability and temperature (De Keersmaecker et al., 2015), (2) compound metrics such as drought indices

(Seddon, Macias-Fauria, Long, Benz, & Willis, 2016), (3) bioclimatic variables (Bruelheide et al., 2018), and (4) combinations of the former (Kling, Auer, Comer, Ackerly, & Hamilton, 2020). To the detriment of research efforts, rarely are all necessary climatic variables for a given study available from a single data product thus necessitating the combination of climate information from several data sources. However, each of the widely-used high resolution climate data sets offer a unique configuration of variables, period covered, methodology and data background, and spatial and temporal resolution, which makes the combination of data from different sources difficult. See Table 1 for an overview of a selection of contemporary climate data sets and their combination of spatial and temporal specifications as well as the number of climatic variables offered by each data product. Accordingly, the study of bioclimatic processes and patterns would be better served by obtaining climate data from a single, internally consistent data source rather than a patchwork of data sets of varying quality and specification.

## *Climate Reanalyses Meet Demands*

Climate reanalysis products represent a major achievement of climate science (Buizza et al., 2018). They meet the demand for high temporal resolutions and abundance of self-consistent climatic information criteria (see Table 1). These products optimally combine a wide range of surface and satellite observations with a dynamical model in order to produce a self-consistent dataset which includes all essential climate variables (ECVs) (Sabater, 2017; Hersbach et al., 2020) effectively eliminating the need for retrieval of data from a multitude of climate products for a full picture of environmental conditions. Reanalyses therefore avoid many of the issues of purely observational products (e.g. WorldClim, CRU). The best reanalyses are often taken as a substitute for observations when studying climate processes and change (Parker, 2016). Two of the most recent, and the most advanced global climate reanalyses have been created by the European Centre for Medium Range Weather Forecasting (ECMWF): ERA5 (Hersbach et al., 2020) and ERA5-Land (Sabater, 2017). The ERA5 reanalysis uses a vast array of observations of the Earth system to constrain a numerical model of the ocean, sea ice, land, and atmosphere using an ensemble data assimilation framework. ERA5 has been demonstrated to improve on data accuracy compared to previously published climate data products (Tang, Qin, Yang, Zhu, & Zhou, 2021). The ERA5 dataset is also the first reanalysis product to make available the uncertainty information of its 10-member ensemble used to create the analysis. This uncertainty is a measure of both the observational uncertainty (which is included in the data assimilation framework) and the stochastic uncertainty. However, this does not account for uncertainty associated with the choice of model physics, which can also be important (Banks et al., 2016). ERA5-Land (Sabater, 2017) is a global land-surface reanalysis that dynamically downscales ERA5 to a resolution of 0.1º (11km). See Table 1. for an overview of ERA5(-Land) data product parameters.

**TABLE 1 - CONTEMPORARY CLIMATE DATA SETS.** A comparison of contemporary high spatial resolution climate data sets which are widely used in analyses of climate impacts. Notes for data availability: [1]… 19 of these are bioclimatic variables which are derivatives of temperature and precipitation data; [2]… 1 of these is elevation data.

| Name | Time-Period | Resolution | | Number of Variables available | Reference |
|---|---|---|---|---|---|
| | | Spatial | Temporal | | |
| WorldClim 2.1 Climatologies | 1960-2018 | 1km | 59 years | 26[1,2] | (Fick & Hijmans, 2017) |
| WorldClim Historical monthly weather data | 1960-2018 | 21km | 1 month | 3 | |
| TerraClimate | 1958-2019 | 16km | 1 month | 14 | (Abatzoglou et al., 2018) |
| CHELSA | 1979-2013 | 1km | 1 month | 46[1] | (Karger et al., 2017) |
| ERA5 | 1950-Today | 30km | 1 hour | 83 | (Hersbach et al., 2020) |

| ERA5-Land | 1981-Today | 11km | 1 hour | 50 | (Sabater, 2017) |

Due to increased data accuracy, temporal resolution, provision of data uncertainty metrics, and number of climate variables provided, ERA5-products are arguably the most appropriate climate data products for macroecological studies.

*LIMITATIONS OF CLIMATE REANALYSIS PRODUCTS*

Despite the advantages of reanalyses, these products have not been widely adopted outside climate science. This is likely a consequence of their relatively coarse spatial resolution (Sabater, 2017; Hersbach et al., 2020). This limitation has motivated several groups to downscale reanalyses to create finer resolution data products (Karger et al., 2017; Abatzoglou et al., 2018). However, none of the existing high-resolution climate products account for the uncertainty in the underlying climate data, or in the downscaling technique effectively biasing user perceptions of their validity in local applications. These products also provide variables which are challenging to robustly statistically downscale to high (~900m) spatial resolution, such as precipitation (Gutmann et al., 2012; Hewitson, Daron, Crane, Zermoglio, & Jack, 2014), or have otherwise violated the assumptions behind statistical downscaling (Chilès & Delfiner, 2012). In addition to spatial resolution mismatches with pre-existing data products, climate reanalysis data can prove challenging to retrieve for potential users. Rather than downloading pre-prepared data files, the user needs to make use of an application programming interface or a webform for retrieval of ERA5(-Land) data. Therefore, to make use of climate reanalysis data effectively, one needs to overcome the two limitations of (1) spatial resolution, and (2) data accessibility.

Here, we present the R package KrigR, which has been developed to address these limitations and create an R-integrated workflow toolbox for handling climate reanalysis data. KrigR can automatically acquire and statistically downscale climate variables using kriging – a Gaussian process regression technique for interpolation (Chilès & Delfiner, 2012). This package can be used to obtain high spatial (~900m) and temporal resolution (hourly) climate data, together with the associated uncertainty.

## 2. STATISTICAL DOWNSCALING

Macroecological studies often make use of climate products at spatial resolutions of 30 arcsecond (~900m). This spatial resolution is roughly one order of magnitude finer than the highest spatial resolution available via ERA5-products (see Table 1). This mismatch of spatial resolutions can be overcome through statistical interpolation methodologies such as Kriging.

### *STATISTICAL INTERPOLATION WITH KRIGING*

Kriging is a two-step process that requires training data that we wish to downscale, and co-variate data both at the resolution of the training data and at our target spatial resolution (Chilès & Delfiner, 2012). In the first step, we fit variograms to our training data and establish covariance functions with our co-variate data at the training resolution. This gives us functions which describe the spatial autocorrelation of our training data, and its relationship with our chosen co-variate(s). During the second step we predict the value of our variable at new locations, in this case at a higher spatial resolution, using co-variate data at the target resolution.

### *ACCURACY OF KRIGING AND IMPLICATIONS FOR BIOLOGICAL STUDIES*

Kriging is a powerful statistical interpolation method capable of accurately interpolating a multitude of climate variables to high spatial resolution with consistent performance across temporal resolutions (Davy & Kusch, 2021). A recent study of vegetation memory patterns across global drylands demonstrated no difference in biological interpretation of spatio-temporal model results when using climate data at native resolution or interpolated from coarser spatial resolution thus proving kriging to be a robust downscaling technique fit for use in biological studies (Kusch, Davy, & Seddon, 2021).

One major advantage to kriging over other statistical interpolation methods is that it preserves the uncertainty obtained when fitting the variogram, which gives us an uncertainty associated with the downscaled data. In KrigR this uncertainty is given as a standard deviation of the uncertainty in the estimate. This statistical uncertainty can be combined with dynamical uncertainty which is calculated by taking the standard deviation of the 10-member ensemble from ERA5 data for a measure of total data uncertainty which can be used to explain differences between climatic data sets (Davy & Kusch, 2021) and should be propagated into biological analyses.

## 3. USING THE KRIGR TOOLBOX

We have prepared a comprehensive overview of how the KrigR package works which can be reached via the KrigR GitHub page.

*WORKFLOW WITH THE KRIGR R-PACKAGE*

The goal of the KrigR package is to make available state-of-the-art climate reanalysis data to R-users at user-specified spatial and temporal resolutions. KrigR does so via two routes summarized in Figure 1.

The first of these is a **Three-Step Process** consisting of (1) obtaining ERA5(-Land) data with calls to the ecmwfr R package (Hufkens, Stauffer, & Campitelli, 2019) and subsequent pre-processing to user specifications of either a rectangular area, a shape (e.g. a country border shapefile), or point-location data. The user specifies the target variable, climate dataset (ERA5 or ERA5-Land), geographic area, time-period and temporal resolution, and optional aggregate metric for the given period (e.g. minimum, maximum, mean, or sum). The second step (2) is obtaining and pre-processing the co-variate data. By default, KrigR provides GMTED2010 (Danielson, J.J., Gesch, 2011) - a digital elevation model (DEM) output – to be used as a co-variate due to the demonstrated close relationship between elevation and a wide range of widely studied climate parameters (Daly, Gibson, Taylor, Johnson, & Pasteris, 2002). The KrigR package downloads the DEM data, masks them to the area/shape/point-location-buffer the user specified and then aggregates the raw DEM data to the user's target resolution and the resolution of the training data. Lastly, KrigR carries out (3) Kriging as made available in R via the automap R package (Hiemstra, Pebesma, Twenhöfel, & Heuvelink, 2009) of the raw ERA5(-Land) data obtained in step 1 using the co-variates obtained in step 2 which results in the output of (A) downscaled ERA5(-Land) data as well as the corresponding (B) statistical uncertainty of the downscaled data. The KrigR package performs additional sanity-checks before Kriging commences, allows for multi-core Kriging, and stores temporary files so that the operation can be terminated and resumed without losing much progress. The user can also specify the degree of localization used in the kriging which affects the estimate, uncertainty, and computational resources used.

By default, the download and pre-processing functions in the KrigR package handle ERA5(-Land) and GMTED2010 data. However, the kriging function of the KrigR package is not limited to the use of these data sets. Third-party data can easily be introduced to the workflow to use the functionality of KrigR on any spatial product with any co-variate as supplied by the user. This is important for two reasons: (1) applicability of kriging has been demonstrated for non-climate spatial products (Bruelheide et al., 2018) and users might also want to downscale other climatic data sets than ERA5(-Land), (2) flexibility in choice of co-variates allows for accurate downscaling of a variety of climate variables and other spatial products. This flexibility in choice of covariates when using KrigR is demonstrated in (Davy & Kusch, 2021). The second route to obtaining high-resolution climate data through KrigR is the **Pipeline**. This involves a single function call that will automatically carry out all three steps explained above. Doing so does not allow for the use of third-party climate products or co-variate data, effectively limiting the user to ERA5(-Land) data and the GMTED2010 co-variate data. Using the Pipeline, a single function call can be used to run the entire process of data downloading, handling, and downscaling.

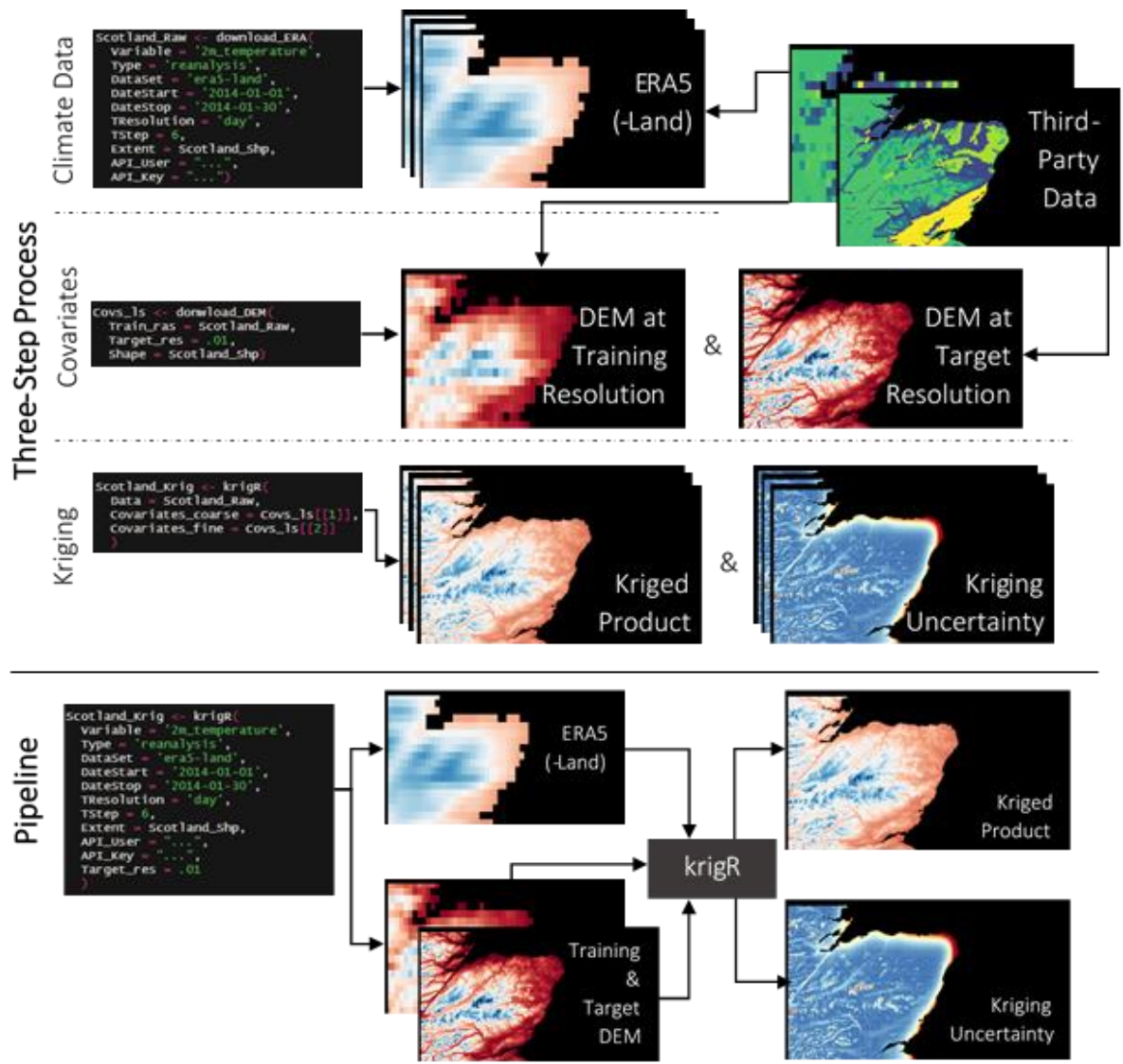

**FIGURE 1 - THE KRIGR WORKFLOW.** KrigR can be executed either through a three-step process which involves three separate function calls to (1) obtain and pre-process climate data, (2) obtain and pre-process co-variate data, and (3) downscale the climate data using the kriging methodology. Alternatively, KrigR can be executed as a pipeline which carries out data download, handling, and downscaling in one function-call. While the three-step process allows for the use of third-party data, the pipeline does not.

*BIOCLIMATIC VARIABLES*

Due to the demonstrated usefulness of bioclimatic variables in biological studies, we have developed the BioClim() function for KrigR which automatically downloads all necessary data and carries out computation of bioclimatic variables as described by Fick & Hijmans, 2017.

The BioClim function can make use of functionality in KrigR thus allowing for (1) limitation of retrieved data to rectangular extents, shapes, or point-location data, (2) multi-core processing of data, (3) storing of temporary files for interruption of the computational process, and (4) full control over where to store temporary files and whether to delete them upon completion of the calculation of bioclimatic variables. Additionally, the BioClim() function allows users to specify for which temporal aggregate to identify extremes thus offering unmatched potential in the quantification of exposure to extreme events. This is of particular importance to bioclimatic variables which record extreme values such as BIO5 and BIO6 (maximum and minimum temperature, respectively) as well as variables reporting climate variability or ranges such as BIO7 (annual temperature range). Finally, the function has been conceptualised in such a way that water availability within the computed bioclimatic variables may be derived from precipitation values (as is the status quo in bioclimatic variables offered by other data products) or any other variable contained within the ERA5(-Land) data products such as soil moisture.

## 4. FUTURE DIRECTIONS

For some variables, such as precipitation, the processes that determine their spatial pattern at finer resolutions than the training data are largely determined by atmospheric dynamics. Therefore, no combination of topographical co-variates is going to enable us to statistically downscale precipitation with high accuracy. We therefore do not recommend statistically downscaling precipitation data. However, there can be alternatives which also tell us about the water availability at high resolution, such as soil moisture, that we can successfully statistically downscale by using the soil properties and topographical properties as co-variates as has been demonstrated in Davy & Kusch, n.d.. To communicate this effectively to users, we are working on a list of recommendations of kriging specifications for individual climate variables to be implemented in the KrigR package in the near future.

Due to high computational costs of kriging at large spatial scales, we are creating high spatial resolution datasets for historical climate, and climate projections for the $21^{st}$ century. For each of these time periods we are preparing a 30 arcsecond (900m) resolution monthly-climatology of surface air temperature for the global land surface. The historical climatology was created using a monthly-climatology created from ERA5-Land for the period 1981-2000 and downscaled using KrigR with elevation as the only covariate.

For the projections for the $21^{st}$ century we make use of the Coupled Model Inter-comparison Project Phase 6 (CMIP6) (Eyring et al., 2016) which includes historical climate simulations and projections for the $21^{st}$ century following shared socioeconomic pathways derived from Integrate Assessment Models (O'Neill et al., 2016). First, we acquire surface air temperature data at monthly resolution from the full set of 36 available CMIP6 models for the historical simulations and the SSP126, SSP245, SSP370, and SSP585 scenarios for the $21^{st}$ century. We then take the ensemble mean of these 36 models for each simulation. Subsequently, we create monthly climatologies for the period 1981-2000 of the historical scenario, and for the periods

2041-2060 and 2081-2100 for the 21$^{st}$ century scenarios. Next, we downscale each of these monthly climatologies to a 30 arcsecond (900m) resolution using elevation as a co-variate. For each of the future periods we then subtract the monthly climatologies from the historical period to create monthly climatologies of temperature anomalies. This was done to remove model biases in regional temperatures. Finally, we added these monthly climatologies of temperature anomalies to the downscaled ERA5-Land monthly climatology for the period 1981-2000. In this way we make use of state-of-the-art CMIP6 projections for temperature changes in the 21$^{st}$ century under multiple realistic scenarios, while retaining the realistic spatial and seasonal variability in temperature obtained from the ERA5-Land reanalysis.

The data products resulting from this work will soon be made available through KrigR.

## 5. CONCLUSIONS

KrigR is a powerful, intuitive, and easy-to-use tool for acquiring and statistically downscaling state-of-the-art climate data. We have integrated the use of the ERA5 family of reanalysis products into KrigR. Currently, these are the most advanced reanalyses. KrigR offers a significant advantage in the field of high-resolution climate datasets by (1) leveraging the important advances behind the creation of the ERA5 reanalyses in terms of observations assimilated, the underlying dynamical model, and the data assimilation methodology; (2) offering access to the full range of essential climate variables from a single, consistent source at high temporal resolution; (3) providing the dynamical and statistical uncertainty associated with the high-resolution data, which allows for uncertainty propagation in downstream modelling efforts as well as a better understanding of data reliability; (4) offers great flexibility to tailor the data and study domain to user needs.

The ability in KrigR to pre-define spatial extent, timescale and period prior to data acquisition helps users overcome an important limitation of conventional workflows. With the rapid growth of climate datasets, the traditional workflow of downloading global data sets and subsequently cropping these to the required areas becomes unmanageable. This data management workflow issue has been an important topic in climate science for a decade (Overpeck, Meehl, Bony, & Easterling, 2011), but is extending to other domains where climate data is used. Thereby, KrigR is a tool with great capability to efficiently provide researchers and other users with climate data tailored to the needs of individual projects while being executed with just a few lines of code in a widely used open-source programming environment.